\documentclass[twocolumn,showpacs,preprintnumbers,amsmath,amssymb,prl,superscriptaddress]{revtex4-1}
\usepackage{bbm}
\usepackage{mathrsfs}
\usepackage{graphicx}
\usepackage{dcolumn}
\usepackage{bm}
\usepackage{amsmath}
\usepackage{amsfonts}
\usepackage{color}
\usepackage[colorlinks=true,citecolor=blue,anchorcolor=blue]{hyperref}
\usepackage{floatrow}

\begin{document}

\title{Topological axion states in magnetic insulator MnBi$_2$Te$_4$ with the quantized magnetoelectric effect}
\author{Dongqin Zhang}
\affiliation{National Laboratory of Solid State Microstructures, School of Physics, Nanjing University, Nanjing 210093, China}
\author{Minji Shi}
\affiliation{National Laboratory of Solid State Microstructures, School of Physics, Nanjing University, Nanjing 210093, China}
\author{Tongshuai Zhu}
\affiliation{National Laboratory of Solid State Microstructures, School of Physics, Nanjing University, Nanjing 210093, China}
\author{Dingyu Xing}
\affiliation{National Laboratory of Solid State Microstructures, School of Physics, Nanjing University, Nanjing 210093, China}
\affiliation{Collaborative Innovation Center of Advanced Microstructures, Nanjing University, Nanjing 210093, China}
\author{Haijun Zhang}
\thanks{zhanghj@nju.edu.cn}
\affiliation{National Laboratory of Solid State Microstructures, School of Physics, Nanjing University, Nanjing 210093, China}
\affiliation{Collaborative Innovation Center of Advanced Microstructures, Nanjing University, Nanjing 210093, China}
\author{Jing Wang}
\thanks{wjingphys@fudan.edu.cn}
\affiliation{State Key Laboratory of Surface Physics, Department of Physics, Fudan University, Shanghai 200433, China}
\affiliation{Collaborative Innovation Center of Advanced Microstructures, Nanjing University, Nanjing 210093, China}
\affiliation{Institute for Nanoelectronic Devices and Quantum Computing, Fudan University, Shanghai 200433, China}

\begin{abstract}
Topological states of quantum matter have attracted great attention in condensed matter physics and materials science. The study of time-reversal-invariant topological states in quantum materials has made tremendous progress. However, the study of magnetic topological states falls much behind due to the complex magnetic structures. Here, we predict the tetradymite-type compound MnBi$_2$Te$_4$ and its related materials host topologically nontrivial magnetic states. The magnetic ground state of MnBi$_2$Te$_4$ is an antiferromagetic topological insulator state with a large topologically non-trivial energy gap ($\sim0.2$~eV). It presents the axion state, which has gapped bulk and surface states, and the quantized topological magnetoelectric effect. The ferromagnetic phase of MnBi$_2$Te$_4$ might lead to a minimal ideal Weyl semimetal.
\end{abstract}

\date{\today}


\maketitle


The discovery of time-reversal-invariant (TRI) topological insulators (TIs)~\cite{hasan2010,qi2011,chiu2016,armitage2018} brings the opportunity to realize a large family of exotic topological phenomena through magnetically gapping the topological surface states (SSs)~\cite{kane2005b,bernevig2006d,koenig2007,qian2014,xia2009,zhang2009,Chen2009,weng2015,huang2015,bernevig2015,xu2015,lv2015,zhu2015,lu2014,ruan2016,bradlyn2016,wang2016,lv2017,zhou2018,wu2018,zhangt2018,tang2018,vergniory2018,chang2013b,mong2010,wan2011,xu2011,tang2016,hua2018,qi2008,wul2016,wang2017c}. Tremendous efforts have been made to introduce magnetism into TRI TIs. One successful example is the first realization of the quantum anomalous Hall (QAH) effect in Cr-doped (Bi,Sb)$_2$Te$_3$ TI thin films~\cite{yu2010,chang2013b,wang2015d}. Aside from the dilute magnetic TIs, intrinsic magnetic materials are expected to provide a clean platform to study magnetic topological states with new interesting topological phenomena. Some magnetic topological states have been theoretically proposed~\cite{macdonald2018}, such as antiferromagnetic (AFM) TI~\cite{mong2010}, dynamical axion field~\cite{li2010}, magnetic Dirac semimetals~\cite{tang2016,jungwirth2017,wang2017b,hua2018} and Weyl semimetals~\cite{wan2011,xu2011,zhang2014a}. Though a few of magnetic Weyl semimetals were experimentally
observed~\cite{kuroda2017}, the study of magnetic topological states falls much behind in experiments due to complex magnetic structure. Therefore, realistic intrinsic magnetic topological materials are highly desired. The class of MnBi$_2$Te$_4$ materials predicted in this Letter provide an ideal platform for emergent magnetic topological phenomena, such as, AFM TI, topological axion state with quantized topological magnetoelectric effect (TME), minimal ideal Weyl semimetal, QAH effect, two-dimensional ferromagnetism and so on.

\begin{figure}
\begin{center}
\includegraphics[width=\textwidth,clip=true]{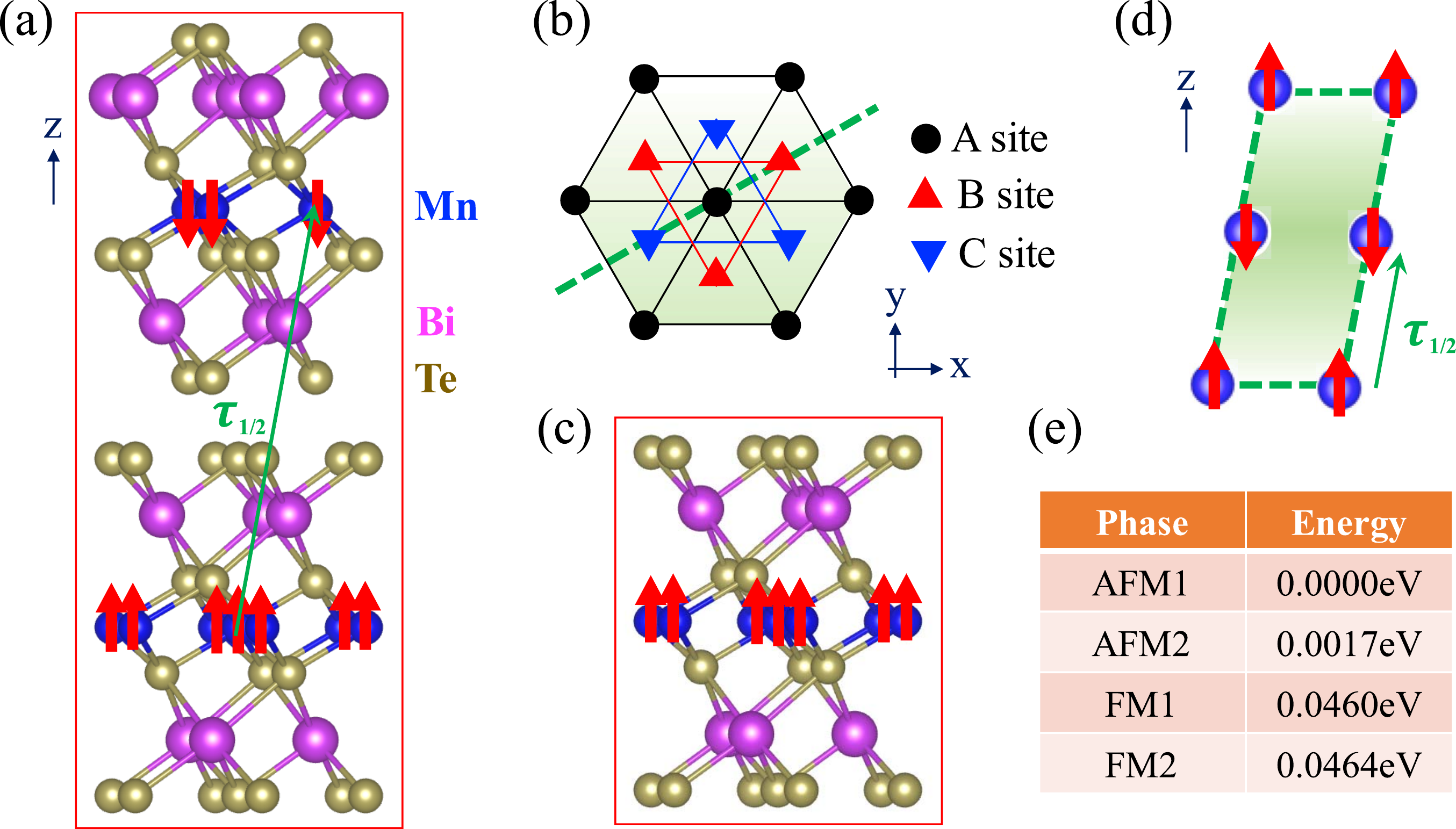}
\end{center}
\caption{Crystal structure and magnetic structure. (a), The unit cell of AFM MnBi$_2$Te$_4$ consists of two SLs. The red arrows represent the spin moment of Mn atom. The green arrow denotes for the half translation operator $\tau_{1/2}$. (b), Schematic top view along the $z$-direction. The triangle lattice in one SL has three different positions, denoted as A, B and C. The dashed green line is used for the (011) plane. (c), The unit cell of FM MnBi$_2$Te$_4$ has one SL. (d), The schematic of the (011) plane, with the blue balls denoting Mn atoms. (e), The calculated total energy for different magnetic ordered states.}
\label{fig1}
\end{figure}

The tetradymite-type compounds XA$_2$B$_4$, also written as XB$\cdot$A$_2$B$_3$ with X $=$ Ge, Sn, Pb or Mn, A $=$ Sb or Bi, and B $=$ Se or Te, crystallize in a rhombohedral crystal structure with the space group $D_{3d}^5$ (No.~166) with seven atoms in one unit cell. We take MnBi$_2$Te$_4$ as an example, which has been successfully synthesized in experiments~\cite{lee2013}. It has layered structures with a triangle lattice, shown in Fig.~\ref{fig1}. The trigonal axis (three-fold rotation symmetry $C_{3z}$) is defined as the $z$ axis,  a binary axis (two-fold rotation symmetry $C_{2x}$) is defined as the $x$ axis and a bisectrix axis (in the reflection plane) is defined as the $y$ axis for the coordinate system. The material consists of seven-atom layers (e.g. Te1-Bi1-Te2-Mn-Te3-Bi2-Te4) arranged along the $z$ direction, known as a septuple layer (SL), which could be simply viewed as the intergrowth of (111) plane of rock-salt structure MnTe within the quintuple layer of TI Bi$_2$Te$_3$ (see Fig.~\ref{fig1}(a) and (c))~\cite{zhang2009}. The coupling between different SLs is the van der Waals type. The existence of inversion symmetry $\mathcal{I}$, with the Mn site as the inversion center, enables us to construct eigenstates with definite parity.

First-principles calculations are employed to investigate the electronic structure of MnBi$_2$Te$_4$, where the detailed methods can be found in the Supplemental Material~\cite{supple}. We find that each Mn atom in MnBi$_2$Te$_4$ tends to have half-filled $d$ orbitals. We performed total energy calculations for different magnetic phases for the three-dimensional MnBi$_2$Te$_4$, and the results are listed in Fig.~\ref{fig1}(e), showing that the A-type AFM phase with the out-of-plane easy axis, denoted as AFM1 (seen in Fig.~\ref{fig1}(a)), is the magnetic ground state. It is ferromagnetic (FM) within the $xy$ plane in each SL, and AFM between neighbor SLs along the $z$ direction, consisting with the previous report~\cite{otrokov2017}. The total energy of the A-type AFM phase AFM2 with the in-plane easy axis is slightly higher than that of AFM1, and much lower than that of FM phase FM1 with the out-of-plane easy axis, which indicates that the magnetic anisotropy is weaker than the effective magnetic exchange interaction between Mn atoms in neighbor SLs. The FM phase FM2 with in-plane easy axis has the highest energy. The Goodenough-Kanamori rule is the key to understand the AFM1 ground state. For the in-plane Mn atomic layer, two nearest Mn atoms are connected through Te atom with the bond `Mn-Te-Mn', whose bonding angle is close to $90$ degree, so the superexchange interaction is expected to induce FM ordering. Contrarily, Mn atoms between neighbor atomic layers are coupled through the bond `Mn-Te-Bi-Te:Te-Bi-Te-Mn', considered as an effective bond `Mn-X-Mn' with a $180$ degree bonding angle, where AFM ordering is induced. In the following discussion, we would focus on the AFM1 (the magnetic ground state) and FM1 (possibly realized through an external magnetic field) states.

\begin{figure}
\begin{center}
\includegraphics[width=\textwidth,clip=true]{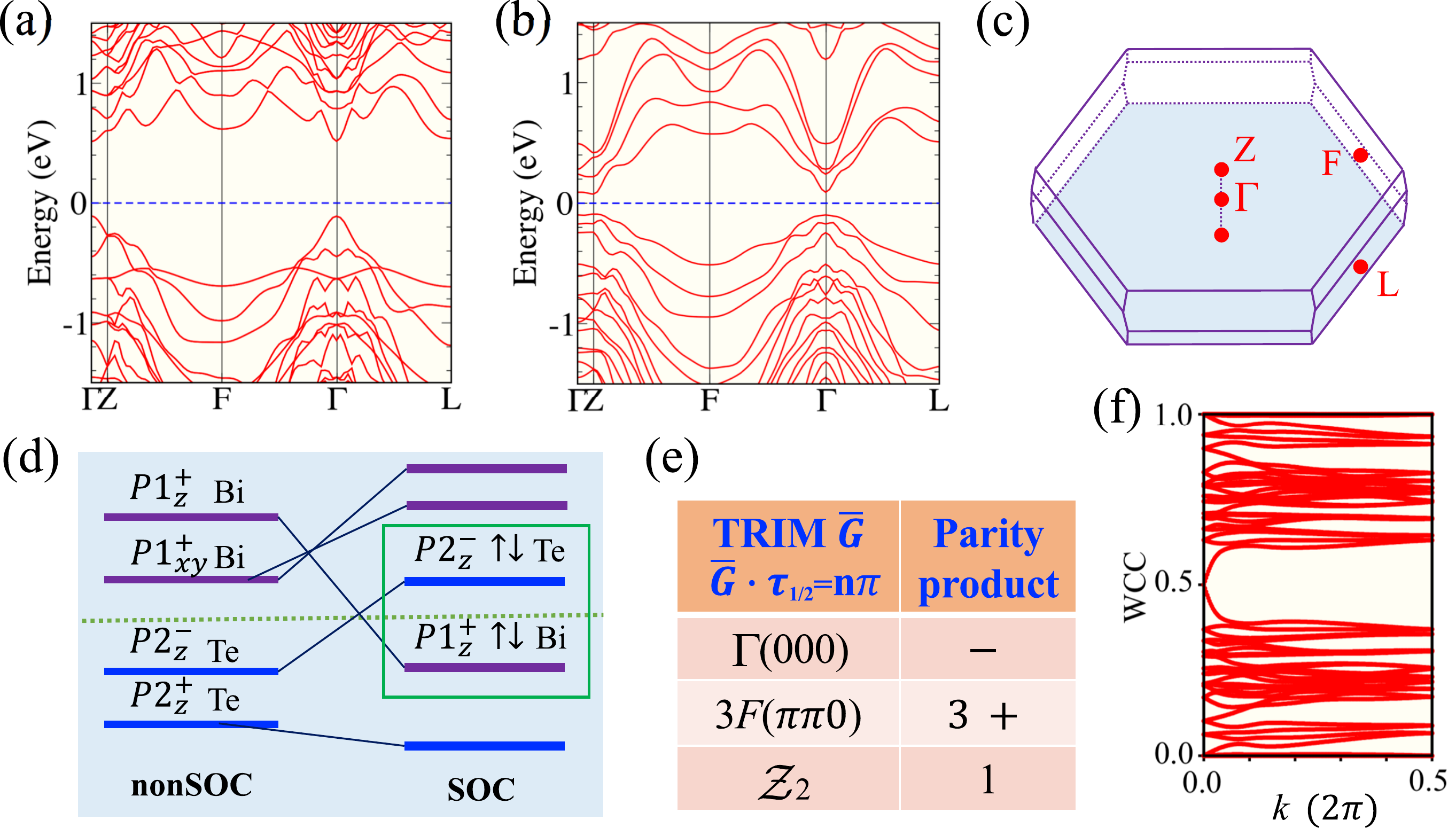}
\end{center}
\caption{Electronic structure of AFM1 MnBi$_2$Te$_4$. (a),(b), The band structure of AFM1 state without (a) and with (b) SOC. (b), The bands are two-fold degenerate due to conserved $\mathcal{I}$ and $\mathcal{S}$. (c), Brillouin zone of MnBi$_2$Te$_4$. The four inequivalent TRIM are $\Gamma(0,0,0)$, $L(\pi,0,0)$, $F(\pi,\pi,0)$ and $Z(\pi,\pi,\pi)$. (d), Schematic diagram of the band inversion at $\Gamma$. The green dotted line represents the Fermi level. (e), The parity product at the TRIM with $\bar{\mathbf{G}}\cdot\tau_{1/2}=n\pi$. (f), The Wannier charge centers (WCC) is calculated in the plane with $\Gamma$ and $3 F$, confirming $\mathcal {Z}_2=1$.}\label{fig2}
\end{figure}

Firstly we investigate the AFM1 ground state. The band structures without and with spin-orbit coupling (SOC) are shown in Fig.~\ref{fig2}(a) and~\ref{fig2}(b), respectively. The time-reversal symmetry $\Theta$ is broken, however, a combined symmetry $\mathcal{S}=\Theta\tau_{1/2}$ is preserved, where $\mathcal{\tau}_{1/2}$ is the half translation operator connecting nearest spin-up and -down Mn atomic layers, marked in Fig.~\ref{fig1}(a). The operator $\mathcal{S}$ is antiunitary with $\mathcal{S}^2=-e^{-i\mathbf{k}\cdot\tau_{1/2}}$.
$\mathcal{S}^2=-1$ on the Brillouin-zone (BZ) plane $\mathbf{k}\cdot\tau_{1/2}=0$. Therefore, similar to $\Theta$ in TRI TI, $\mathcal{S}$ could also lead to a $\mathcal{Z}_2$ classification~\cite{mong2010}, where the topological invariant is well defined on the BZ plane with $\mathbf{k}\cdot\tau_{1/2}=0$. One can see an anti-crossing feature around the $\Gamma$ point from the band inversion, suggesting that MnBi$_2$Te$_4$ might be topologically nontrivial. Since $\mathcal{I}$ is still preserved, the $\mathcal{Z}_2$ invariant is simply determined by the parity of the wave functions at TRI momenta (TRIM) in the Brillouin zone~\cite{fu2007a}. Here we only need consider the four TRIM ($\Gamma$ and three $F$) with $\bar{\mathbf{G}}\cdot\tau_{1/2}=n\pi$. As expected, by turning on SOC, the parity of one occupied band is changed at $\Gamma$ point from band inversion between the $|P1_z^+\rangle$ of Bi and the $|P2_z^-\rangle$ of Te, schematically shown in Fig.~\ref{fig2}(d), whereas the parity remains unchanged for all occupied bands at the other three momenta $F$ (see Fig.~\ref{fig2}(e)), so $\mathcal{Z}_2=1$. We also employ the Willson loop method~\cite{yu2011} to confirm the $\mathcal{Z}_2$ invariant in Fig.~\ref{fig2}(f), concluding that AFM MnBi$_2$Te$_4$ is an AFM TI. Especially, we notice that a large energy gap of about $0.2$~eV is obtained in Fig.~\ref{fig2}(b).

The existence of topological SSs is one of the most important properties of TIs. However, the TI state in AFM MnBi$_2$Te$_4$ protected by $\mathcal{S}$ is topological in a weaker sense than the strong TI protected by $\Theta$, which manifests in that the existence of gapless SS depends on the surface plane. As shown in Fig.~\ref{fig4}(a) and \ref{fig4}(c), there is gapped SSs on the (111) surface accompanied by a triangular Fermi surface, for $\mathcal{S}$ is broken. As shown in Fig.~\ref{fig4}(b), only on the $\mathcal{S}$-preserving surfaces such as (011) surface, the gapless SSs are topologically protected which forms a single Dirac-cone-type dispersion at $\Gamma$.

\begin{figure}
\begin{center}
\includegraphics[width=\textwidth,clip=true]{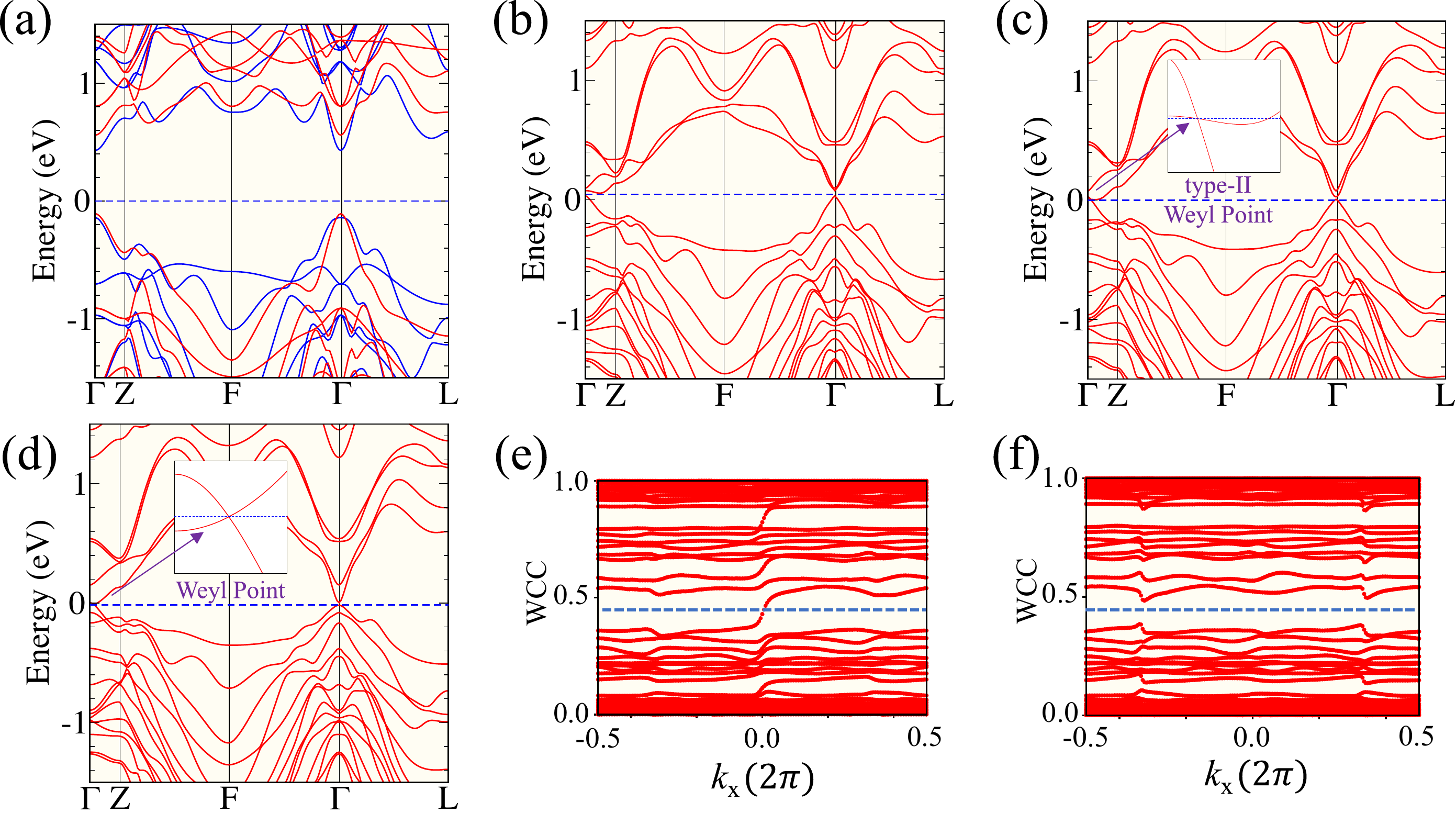}
\end{center}
\caption{Electronic structure of FM1 MnBi$_2$Te$_4$. (a), Band structure for FM1 state without SOC. The dashed line indicates the Fermi level. The red/blue lines are spin-up/-down bands. (b)-(d), Band structures for FM1 state with SOC are calculated by the LDA+U ($U=3$~eV) functional with experimental lattice constants ($a_0,c_0$) in (b), extended lattice constants ($1.005a_0,1.005c_0$) in (c) and ($1.01a_0,1.01c_0$) in (d), respectively. The system has the transition from FM insulator to type-II Weyl semimetal, and finally to ideal Weyl semimetal. (e),(f), The evolution of WCC along the $k_x$ direction in the $k_z=0$ plane (e) and in the $k_z=\pi$ plane (f). The WCCs cross the reference horizontal line once in (e), indicating the Chern number $C=1$ in the $k_z=0$ plane. Oppositely, the WCCs don't cross the reference line in (f), indicating the Chern number $C=0$ in the $k_z=\pi$ plane.}
\label{fig3}
\end{figure}

For the FM1 state of MnBi$_2$Te$_4$, the band structures without and with SOC effect are shown in Fig.~\ref{fig3}. MnBi$_2$Te$_4$ is a FM insulator with the experimental lattice constant ($a_0,c_0$), shown in Fig.~\ref{fig3}(b). Interestingly, we find that the band structure is sensitive to the lattice constant. When the lattice constant is slightly extended, it first becomes a type-II Weyl semimetal with ($1.005a_0,1.005c_0$) and then becomes a minimal ideal Weyl semimetal with ($1.01a_0,1.01c_0$), hosting two Weyl points at the Fermi level without other bulk bands mixing, shown in Fig.~\ref{fig3}(c)-(d). The Willson loop calculations, shown in Fig.~\ref{fig3}(e) and \ref{fig3}(f), suggest that the Chern number $C=1$ at $k_z=0$ plane, and $C=0$ at $k_z=\pi$ plane, which is consist with the ideal Weyl semimetal in Fig.~\ref{fig3}(d). Furthermore, the SSs of FM1 state on different typical surfaces are calculated. In Fig.~\ref{fig4}(d), bulk states projected on the (111) surface have no energy gap, for the two Weyl points are exactly projected to the surface $\bar{\Gamma}$ point. In Fig.~\ref{fig4}(e) and \ref{fig4}(f), one can clearly see the surface Fermi arcs connecting to the two ideal Weyl points are separated ($\sim0.06$~\AA$^{-1}$).

\emph{Low-energy effective model.} As the topological nature is determined by the physics near the $\Gamma$ point, a simple effective Hamiltonian can be written down to characterize the low-energy long-wavelength properties of the system. We start from the four low-lying states $|P1_z^+,\uparrow(\downarrow)\rangle$ and $|P2_z^-,\uparrow(\downarrow)\rangle$ at the $\Gamma$ point. Here the superscripts `$+$', `$-$' stand for the parity of the corresponding states. Without the SOC effect, around the Fermi energy, the bonding state $|P1_z^+\rangle$ of two Bi layers stays above of the antibonding state $|P2_z^-\rangle$ of two Te layers (Te1 and Te4 in SLs). As shown in Fig.~\ref{fig2}(d), the SOC mixes spin and orbital angular momenta while preserving the total angular momentum, and $|P1_z^+,\uparrow(\downarrow)\rangle$ state is pushed down and the $|P2_z^-,\uparrow(\downarrow)\rangle$ state is pushed up, leading to the band inversion and parity exchange. In the nonmagnetic state, the symmetries of the system are $\Theta$, $\mathcal{I}$, $C_{3z}$ and $C_{2x}$. In the basis of ($|P1_z^+,\uparrow\rangle$, $|P2_z^-,\uparrow\rangle$, $|P1_z^+,\downarrow\rangle$, $|P2_z^-,\downarrow\rangle$), the representation of symmetry operations is given by $\Theta=1_{2\times2}\otimes i\sigma^y\mathcal{K}$, $\mathcal{I}=\tau^z\otimes1_{2\times2}$,  $C_{3z}=\exp(1_{2\times2}\otimes  i(\pi/3)\sigma^z))$ and $C_{2x}=\exp(\tau^z\otimes  i(\pi/2)\sigma^x))$, where $\mathcal{K}$ is the complex conjugation operator, $\sigma^{x,y,z}$ and $\tau^{x,y,z}$ denote the Pauli matrices in the spin and orbital space, respectively. By requiring these four symmetries and keeping only the terms up to quadratic order in $\mathbf{k}$, we obtain the following generic form of the effective Hamiltonian for nonmagnetic state
\begin{equation}
\mathcal{H}_\text{N}(\mathbf k) = \epsilon_0(\mathbf k)+\begin{pmatrix}
M_\gamma(\mathbf k) & A_1k_z &0 & A_2k_-\\
A_1k_z & -M_\gamma(\mathbf k) & A_2k_- & 0\\
0 & A_2k_+ & M_\gamma(\mathbf k) & -A_1k_z\\
A_2 k_+ & 0& -A_1k_z & -M_\gamma(\mathbf k)
\end{pmatrix},
\nonumber
\end{equation}
where $k_\pm=k_x\pm ik_y$, $\epsilon_0(\mathbf k)=C+D_1k_z^2+D_2(k_x^2+k_y^2)$ and $M_\gamma(\mathbf k)=M^{\gamma}_{0}+B^{\gamma}_{1}k_z^2+B^{\gamma}_{2}(k_x^2+k_y^2)$.

The FM1 state breaks $\Theta$ and $C_{2x}$ but preserves the combined $C_{2x}\Theta$, therefore the effective Hamiltonian for FM1 is obtained by adding perturbative term $\delta\mathcal{H}_\text{FM1}(\mathbf k)$ respecting the corresponding symmetries into $\mathcal{H}_{\text{N}}(\mathbf{k})$, which is
\begin{equation}
\delta\mathcal{H}_\text{FM1}(\mathbf k) =
\begin{pmatrix}
M_1(\mathbf{k}) & A_3k_z & 0 & A_4k_-\\
A_3k_z & M_2(\mathbf{k}) & -A_4k_-&0\\
0 & -A_4 k_+ & -M_1(\mathbf{k}) & A_3k_z\\
A_4 k_+ & 0 & A_3k_z & -M_2(\mathbf{k})
\end{pmatrix},
\nonumber
\end{equation}
where $M_{1,2}(\mathbf{k})=M_\alpha(\mathbf{k})\pm M_{\beta}(\mathbf{k})$, and $M_j(\mathbf k) = M^{j}_{0} + B^j_{1}k_z^2+B^j_{2}(k_x^2+k_y^2)$ with $j = \alpha, \beta$. By fitting the energy spectrum of the effective Hamiltonian with that of the first-principles calculation, the parameters in the effective model can be determined, which can be found in the Supplemental Material~\cite{supple}. The $M_{1,2}$ terms characterize the Zeeman coupling with the magnetized Mn orbitals, and in general $M_1\neq M_2$ denotes the different effective $g$-factor of $|P1_z^+,\uparrow(\downarrow)\rangle$ and $|P2_z^-,\uparrow(\downarrow)\rangle$.

The AFM1 state breaks $\Theta$ but preserves $\mathcal{S}=1_{2\times2}\otimes i\sigma^y\mathcal{K}e^{i\mathbf{k}\cdot\tau_{1/2}}$, and the unit cell doubles compared to FM1 state. For simplicity, we obtain the low-energy four-band model similar to the above analysis. From band structure analysis, the four bands close the Fermi energy in the AFM1 state are the new bonding state $|P1_z^{\prime+},\uparrow(\downarrow)\rangle$ of four Bi layers and the antibonding state $|P2_z^{\prime-},\uparrow(\downarrow)\rangle$ of four Te layers (two Te1 and two Te4 in neighboring SLs). In the basis of ($|P1_z^{\prime+},\uparrow\rangle$, $|P2_z^{\prime-},\uparrow\rangle$, $|P1_z^{\prime+},\downarrow\rangle$, $|P2_z^{\prime-},\downarrow\rangle$), by requiring the symmetries $\mathcal{I}$, $C_{3z}$ and $\mathcal{S}$,  we get the effective Hamiltonian for AFM1 which has the \emph{same} expression as $\mathcal{H}_{\text{N}}(\mathbf{k})$ but with different parameters. The AFM1 model and fitting parameters are listed in the Supplemental Material~\cite{supple}.

\begin{figure}
\begin{center}
\includegraphics[width=\textwidth,clip=true]{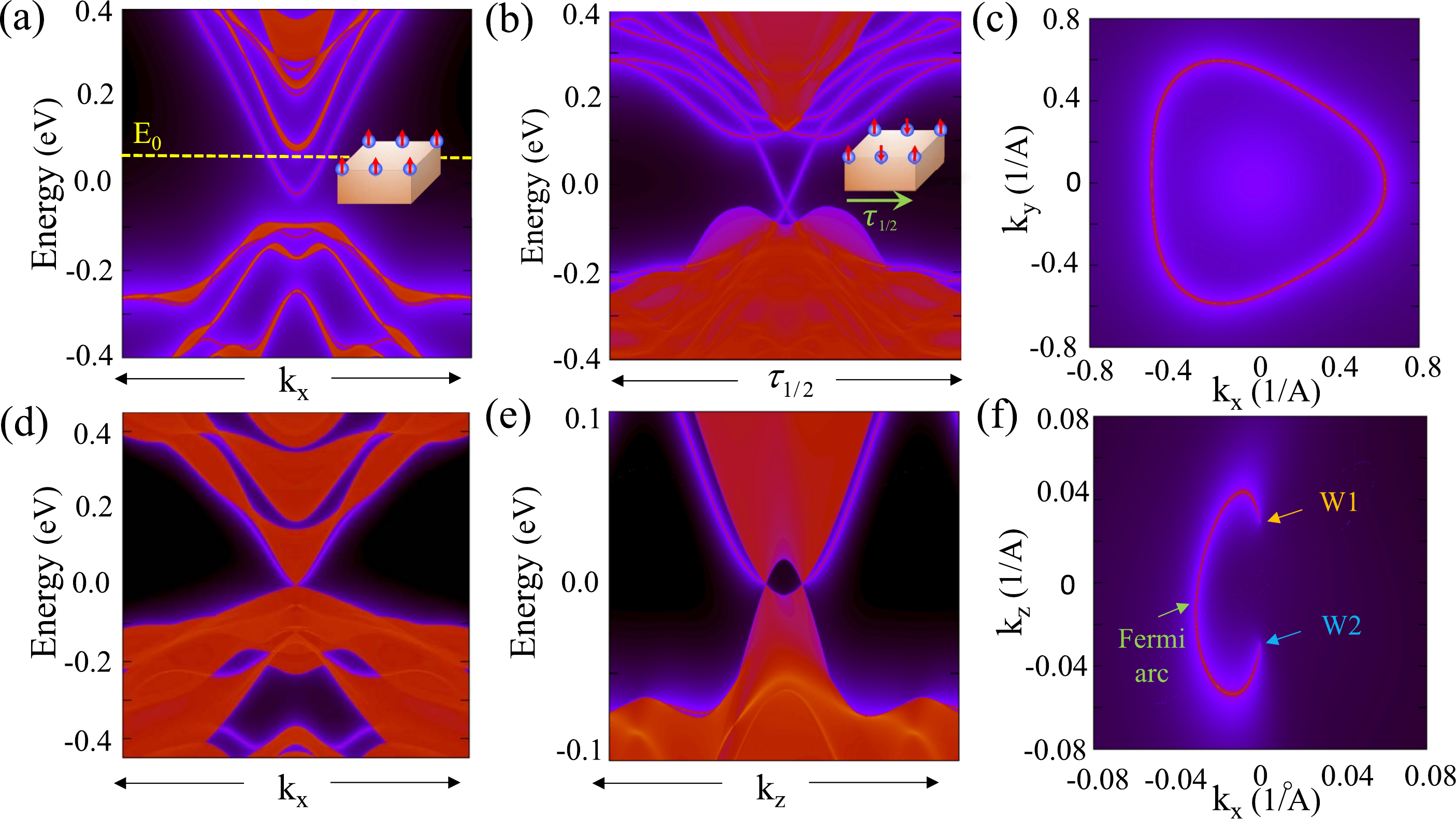}
\end{center}
\caption{Surface states. (a),(b), Energy and momentum dependence of the local density of states (LDOS) for AFM1 phase on the (111) and (011) surfaces, respectively. In (a), The SSs on (111) surface are fully gapped due to the $\mathcal{S}$ symmetry broken. In (b), The gapless SSs can be seen at the $\Gamma$ point with a linear dispersion in the bulk gap on the $\mathcal{S}$-preserving (011) surface. (c), Fermi surface on the (111) surface at the energy level $E_0$ in (a) presents the triangle shape, different from the hexagonal shape in TI Bi$_2$Se$_3$. (d),(e), Energy and momentum dependence of the LDOS for FM1 phase on the (111) and (011) surfaces, respectively. In (e), the two Weyl points are seen along the $k_z$ direction. (f), There are two Fermi arc connecting the Weyl points W1 and W2, indicating the ideal Weyl semimetal feature.} \label{fig4}
\end{figure}

\emph{Axion state and topological response.} The topological electromagnetic response from AFM TI is described by the topological $\theta$ term, $S_{\theta}=(\theta/2\pi)(\alpha/2\pi)\int d^3 xdt \mathbf{E}\cdot \mathbf{B}$. Here, $\mathbf{E}$ and $\mathbf{B}$ are the conventional electromagnetic fields inside the insulator, $\alpha=e^2/\hbar c$ is the fine-structure constant, $e$ is electron charge, and $\theta$ is dimensionless pseudoscalar parameter and defined only modulo $2\pi$. Physically, $\theta$ has an explicit microscopic expression of the momentum space Chern-Simons form~\cite{qi2008,essin2009}. $\mathcal{S}$ constrains $\theta$ to be quantized, namely $\theta=-\theta+2\pi n$ for integer $n$, thus $\theta=\pi$ for AFM TI. From the effective action with open boundary conditions, we know that $\theta=\pi$ implies a surface quantized Hall conductance of $\sigma_{xy}=e^2/2h$. This half quantized Hall effect on the surface is the physical origin behind the topological TME effect. For a finite TRI TI, $\Theta$ forces TME to vanish, where the surface and bulk states contribution to TME precisely cancel each other~\cite{qi2008,mulligan2013,rosenow2017}. As is suggested in Ref.~\cite{qi2008,wang2015b}, to obtain the quantized TME in TIs, one must fulfill the following stringent requirements. First, all surfaces are gapped by magnetic ordering. Second, the Fermi level is finely tuned into the magnetically induced surface gap to keep the bulk truly insulating. Third, the film of TI material should be thick enough to eliminate the finite-size effect. However, the previous proposals on TME in the FM-TI heterostructure have several drawbacks. First, the gapless SSs on side surfaces are hard to eliminate~\cite{wang2015b,mogi2017,xiao2018}, which will destroy the TME. Second, the surface gap is tiny due to weak magnetic proximity effect.

Interestingly, MnBi$_2$Te$_4$ provides a feasible platform for quantized TME, which has not been
experimentally observed. One advantage is that the $\mathcal{S}$ breaking surfaces are gapped by material's own time-reversal breaking, thus allowing a non-vanishing TME. One can simply grow realistic materials without any $\mathcal{S}$-preserving surfaces or apply a small in plane magnetic field. Such axion state has fully gapped bulk and surfaces, and the same topological response as AFM TI with $\theta=\pi$. The second advantage is that the surface gap induced by intrinsic magnetism is large of about $0.1$~eV. Furthermore, the finite-size effect is negligible when the film exceed 4 SLs~\cite{supple}. Experimentally, such quantized TME can be observed by measuring the induction of a parallel polarization current when an ac magnetic field is applied~\cite{wang2015b}, which is $\mathcal{J}=(\theta/\pi)(e^2/2h)(\partial B_x/\partial t)\ell d$. Here, $d$ and $\ell$ are the thickness and width of the MnBi$_2$Te$_4$ sample. For an estimation, taking $B_x=B_0e^{-i\omega t}$, $B_0=10$~G, $\omega/2\pi=1$~GHz, $d=50$~nm, $\theta=\pi$, and $\ell=400$~$\mu$m, we have $\mathcal{J}=-i\mathcal{J}_0e^{-i\omega t}$ with $\mathcal{J}_0=2.22$~nA, in the range accessible by experiments.

It is worth mentioning that the N\'{e}el order in AFM1 state is essentially different from that in dynamical axion field proposed in Ref.~\cite{li2010}. In the latter case, the N\'{e}el order breaks $\Theta$ and $\mathcal{I}$, but conserves $\mathcal{I}\Theta$. The magnetic fluctuation of the N\'{e}el order leads to linear contribution to the fluctuation of axion field, and the static $\theta$ deviates from $\pi$. While in the case of AFM1 MnBi$_2$Te$_4$, the N\'{e}el order conserves both $\mathcal{I}$ and $\mathcal{S}$, thus the static $\theta=\pi$, and to the linear order, the magnetic fluctuation has no contribution to the dynamics of axion field~\cite{li2010,taguchi2018}.

\emph{Materials.} Other tetradymite-type compounds XBi$_2$Te$_4$, XBi$_2$Se$_4$ and XSb$_2$Te$_4$ (X = Mn or Eu), if with the same rhombohedral crystal structure, are also promising candidates to host magnetic topological states similar to MnBi$_2$Te$_4$. For example, EuBi$_2$Te$_4$ is another AFM TI, and MnSb$_2$Te$_4$ is at the topological quantum critical point~\cite{supple}. Actually, tetradymite-type compounds XB$\cdot$A$_2$B$_3$  belong to a large class of ternary chalcogenides materials (XB)$_n$$\cdot($A$_2$B$_3$)$_m$ with X = (Ge, Sn or Pb), A = (Sb or Bi) and B = (Se or Te), most of which were found to be TIs~\cite{jin2011}. Interestingly, (GeTe)$_n$(Sb$_2$Te$_3$)$_m$ and (GeTe)$_n$(Bi$_2$Te$_3$)$_m$ have been widely studied as phase change memory materials~\cite{wuttig2007}. By tuning the layer index $m$ and $n$, we can play with the crystal structure, the topological property, and the magnetic property of the series of materials (XB)$_n$$\cdot($A$_2$B$_3$)$_m$, which opens a broad way to study emergent phenomena of magnetic topological states. For example, the dynamic axion field may be obtained in these systems.

Finally, the intrinsic magnetism and band inversion further lead to QAH effect in odd layer MnBi$_2$Te$_4$ thin film with $\mathcal{I}\Theta$ broken~\cite{supple}. The intrinsic magnetic topological materials predicted here are simple and easy to control, which could host extremely rich topological quantum states in different spatial dimensions and are promising for investigating other exotic emergent particles such as Majorana fermions.

\begin{acknowledgments}
We thank Ke He for stimulating discussions. H.Z. is supported by the Natural Science Foundation of China (Grant Nos.~11674165, 11834006), the Fok Ying-Tong Education Foundation of China (Grant No.~161006). J.W. is supported by the Natural Science Foundation of China through Grant No.~11774065, the National Key Research Program of China under Grant No.~2016YFA0300703, the Natural Science Foundation of Shanghai under Grant No.~17ZR1442500, the Open Research Fund Program of the State Key Laboratory of Low-Dimensional Quantum Physics through Contract No.~KF201606, and by Fudan University Initiative Scientific Research Program.

D.Z. and M.S. contributed equally to this work.
\end{acknowledgments}

\emph{Note added}: Recently, we learned of the experimental papers in the same material by Gong \emph{et al}~\cite{gong2018} and Otrokov \emph{et al}~\cite{otrokov2018}.

\end{document}